\documentclass[aps,prl,floatfix,preprint]{revtex4}
\usepackage[pdftex,dvipdfm,dvips]{graphicx}
\usepackage{amstext}
\usepackage{amsthm}
\usepackage{amssymb, amsmath}
\usepackage{subfigure}
\usepackage{natbib}
\usepackage{float}
\usepackage{url}
\usepackage[hidelinks]{hyperref}
\usepackage{doi}

\begin{document}
\title{\textbf{\large{Non-monotonic resistance noise in the charge density wave pinned state in single nanoribbons of CDW conductor NbSe$_{3}$}}}
\author{Zhenzhong Shi$^{1}$}
\author{Peter M Marley$^{2}$}
\author{Katie Farley$^{2,3}$}
\author{Sarbajit Banerjee$^{2,3}$}
\author{G. Sambandamurthy$^{1}$}
\affiliation{Department of Physics$^{1}$}
\affiliation{Department of Chemistry$^{2}$, The State University of New York at Buffalo, Buffalo, NY 14260, USA}
\affiliation{Department of Chemistry$^{3}$, Texas A \& M University, College Station, TX 77843, USA}

\begin{abstract}

Electrical transport and broadband resistance noise measurements in an ultra low frequency window (30 mHz - 8 Hz) are carried out in single nanoribbon devices of charge density wave (CDW) conductor NbSe$_{3}$. In the temperature and electric field range where the CDW is expected to be completed pinned by residual impurities, a hitherto unseen non-monotonic behavior in the noise magnitude vs. electric field is observed. This behavior can be attributed to the proliferation of thermally activated phase slip events and this idea is supported by the observation of a smeared activated behavior described by the Dutta-Horn relation. Certain features of the temperature dependence of the noise magnitude do not follow an activated behavior pointing to a complex origin of the fluctuations in a CDW system.

\end{abstract}

\maketitle

The interaction between the elasticity of matter and various randomly distributed impurities plays an important role in several physical systems, such as Wigner crystals \cite{Eguiluz1983,Tanatar1989}, vortices in type II superconductors \cite{Tinkham1996} and charge density wave (CDW) materials \cite{Gruner1988,Monceau2012}. In these systems, the long range order could be destroyed by the pinning forces of topological defects. However, an externally applied driving force can overcome the pinning strength and restore the long range order of the pinned objects \cite{Eguiluz1983,Tanatar1989,Tinkham1996,Gruner1988,Monceau2012}. Well studied examples of these phenomena are materials exhibiting CDW properties. A variety of experimental techniques, such as X-ray diffraction \cite{DiCarlo1993,Isakovic2006,Pinsolle2012}, scanning tunneling microscopy (STM) \cite{Coleman1985,Brun2009}, electrical transport measurements \cite{MonAeau1976,Thompson1981,Monceau2012} and broadband noise (BBN) measurements \cite{Bhattacharya1985,Bloom1993,Marley1992} have been used to study the physics of CDW materials. Each technique probes the system from a different perspective: diffraction and transport techniques probe averaged effects over a certain size of the sample, and the STM technique is mainly used to probe the surface of the sample. The noise measurement has its advantage in probing the microscopic dynamics of the systems, such as domain wall dynamics in magnetic systems \cite{PRLv105p067206} and the glassy dynamics in the two dimensional electron systems \cite{PRLv88p236401,PRLv89p276401,Koushik2011} and in high T$ _{C} $ superconductors \cite{Raicevic2008}. 

In CDW systems, the noise measurement is a valuable tool thanks to its sensitivity to the dynamics of the deformed CDWs at pinning sites \cite{RMPv60p537,COISSMSv6p67}. Previous studies of the noise in the CDW system have provided a good knowledge of the microscopic dynamics of the CDW depinning process. These studies mostly focused on the CDW depinning and sliding regime. The origin of the noise in these regimes has been explained by two main concepts. The first hypothesis by
Bhattacharya \textit{et al.} \cite{Bhattacharya1985, Bhattacharya1989} considers mainly the quasi-equilibrium fluctuations among metastable states with different pinning forces. On the other hand, Bloom \textit{et al.} \cite{Bloom1994} included the non-equilibrium drag force of the flowing CDW to understand the physical properties in the sliding regime. Though CDW phase creep and the CDW phase slippage have been shown to be important in this pinned regime \cite{Zaitsev-Zotov1993,Lemay1999,Pinsolle2012}, a direct measurement of resistance noise in the CDW pinned regime has not been performed. A study of resistance noise in this regime is particularly important because it could potentially provide information about the above mentioned microscopic processes. Only a study that indirectly probed the conductance fluctuation in the pinned regime was done previously, where the authors measured the transition probability of a two state random telegraph signal as a function of time during which the system spent in the pinned state \cite{Bloom1993}. 

A bottleneck in the previous studies of the resistance noise in the CDW pinned regime where phase slip/creep is expected to occur is the ability to detect extremely small signals. This is due to two reasons: first, the studied samples were large in dimension ($\sim$mm $\times \sim\mu$m$^{2}$) where single phase slip events may not be easily observable when the measurements are averaged over the entire sample. Second, the frequency window ($\sim$1 Hz - $\sim$1 MHz) that was used in previous measurements \cite{Maher1991,Bloom1993} was probably too high to observe the slow dynamics in the pinned regime. Both shortcomings can be overcome by reducing the sample dimension and through the selection of an appropriately low frequency window. In this article, we examine the hitherto unexplored slow dynamics manifested in the pinned regime of CDW systems, thereby directly exploring the atomistic motion of the periodic lattice in this regime. We report for the first time the 1/f noise measurement in a frequency window (30 mHz - 8 Hz) in the CDW pinned state of sub-micron sized NbSe$_{3}$ nanoribbons which shows hitherto unseen features, such as a nonmonotonic behavior of electric field dependence of the noise magnitude in the pinned CDW state.

\begin{figure*}
\includegraphics[width=6in,height=5in]{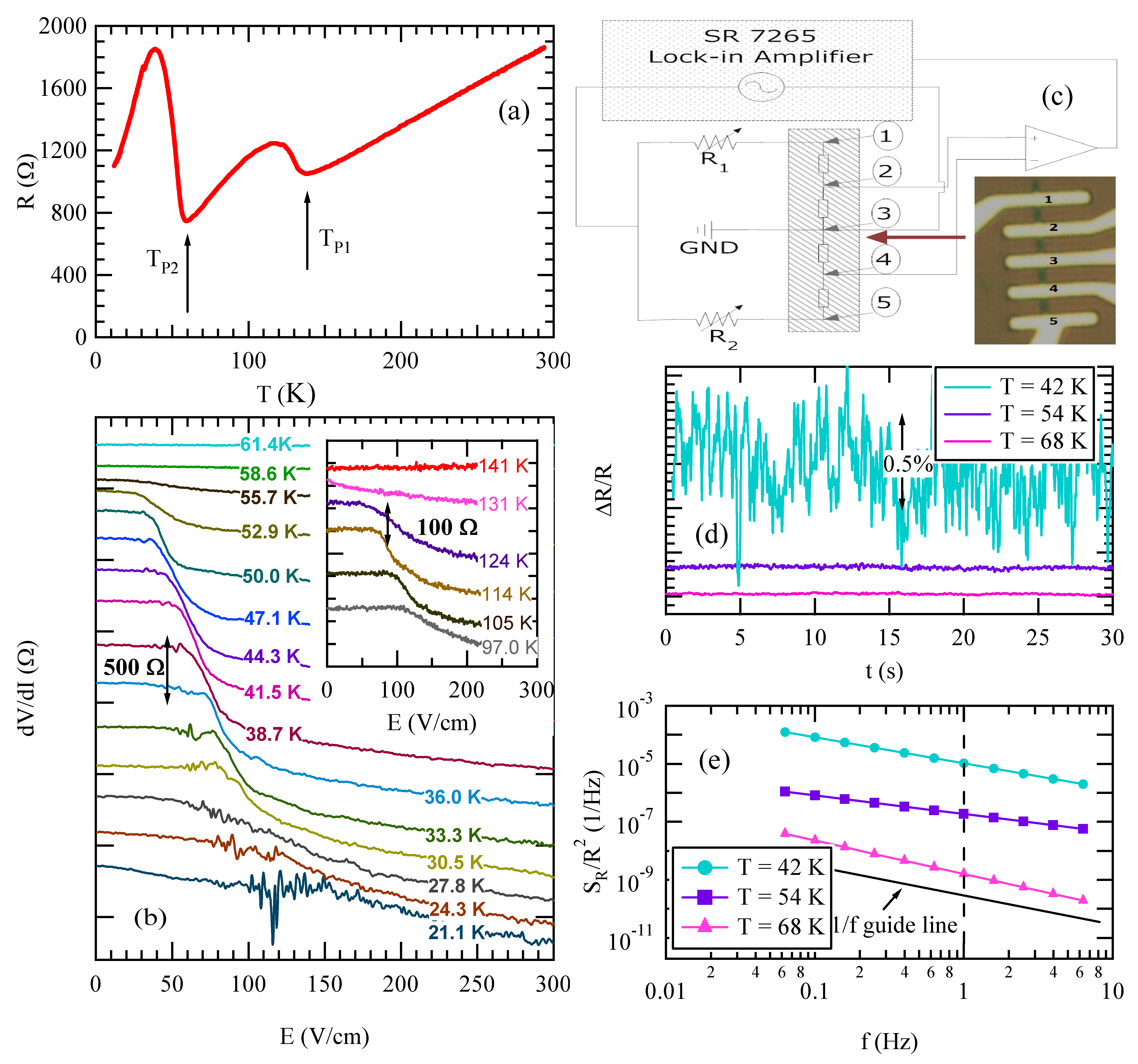}
\caption{(a) Resistance vs. Temperature of a typical NbSe$_{3}$ nanoribbon device (device A), measured with the standard four probe method. T$_{P1}$ is 139 K and T$_{P2}$ is 59 K. (b) Differential resistance as a function of electric field around T$_{P2}$ (main plot) and T$_{P1}$ (insert), measured from device A. A threshold behavior can be seen at temperatures below T$_{P1}$ and T$_{P2}$. (c) A Schematic representation of the Wheatstone bridge based noise measurement set up. The voltage fluctuation ${\delta}v(t) = \langle{v-\langle{v}\rangle}\rangle$, across the current biased resistor is measured. The insert is an optical micrograph of a typical 5-terminal device. The contacts are 5 $\mu$m wide and are spaced 5 $\mu$m apart. An ac current is sent from contact 1 and contact 5 to the ground contact 3, and the voltage fluctuation ($\delta$V) is measured across contact 2 and contact 4. (d) The resistance fluctuations from a NbSe$_{3}$ nanoribbon device are shown in the time domain at three representative temperatures (T = 42 K, 54 K and 68 K ) across  T$_{P2}$ (e) The normalized power spectral density (PSD) of the resistance calculated from the three traces in (d), shown by a fit and representative data points. They all follow a generic 1/f$ ^{\alpha} $ behavior. The frequency exponent $ \alpha $ is temperature dependent and varies around 1. For T = 42 K, 54 K and 68 K, $ \alpha $ is 0.90, 0.64, and 1.15 respectively.}
\label{Figure1.png}
\end{figure*}

We fabricated multi-terminal electrical devices from individual nanoribbons of single crystalline NbSe$_{3}$ using standard optical and/or electron beam lithography and metallization techniques. The growth and structural characterization of the nanowires were reported elsewhere \cite{Stabile2011} and the typical dimensions of the nanoribbons used in this study were 50 $\mu$m $\times$ 500 nm $\times$ 100 nm. Typically five gold contacts, which are 5 $\mu$m wide with a separation of 5 $\mu$m, were evaporated on top of a single nanoribbon thereby making good Ohmic contacts (see figure 1(c) for a typical device).  Transport measurements were carried out in a liquid Helium Variable Temperature Insert (VTI) between 15 K and 300 K,  using standard ac/dc techniques with lock-in amplifiers and dc sources.

The temperature ($T$) dependence of resistance for a typical single nanoribbon device (Device A) is shown in Fig.  \ref{Figure1.png} (a), where two Peierls transitions can be seen at T$_{P1}$ = 139 K and T$_{P2}$ = 59 K, as expected. Sample to sample variations in T$_{P1}$ and T$_{P2}$ values lie within a few degrees in the nine devices that were measured during this study. In figure 1(b), differential resistance traces of the sample measured at several temperatures around T$_{P1}$ and T$_{P2}$ are plotted as a function of electric field ($E$) across the sample. A clear threshold behavior can be seen at temperatures where a pinned CDW state is expected and the increasing electric field depins the CDW state \cite{MonAeau1976}. A low residual resistance ratio (R$ _{300K} $/R$ _{15K} $ = 1.5 - 8) and large threshold field (10 V/cm - 100 V/cm) for CDW depinning  found in our samples suggest strong finite size effects, and are consistent with recent transport studies on NbSe$_{3}$ nanoribbons of comparable sizes \cite{Thorne1992,Slot2004,nl0480722}.

In order to obtain the $T$ and $E$ dependence of noise behavior, time-dependent voltage measurements across the samples were performed using an improved Wheatstone bridge configuration under balance condition \cite{RSIv58p985,GhoshThesis}, as shown in  Fig. \ref{Figure1.png} (c). The entire setup was carefully grounded and shielded to minimize the background noise arising from external sources. An ac excitation voltage, provided by a SR7265 lock-in amplifier, was applied across the Wheatstone bridge. The frequency of the ac voltage was set at 274 Hz, which was well above our measured frequency window (30 mHz to 8 Hz) and much smaller than CDW depinning frequency in NbSe$_{3}$ (10$ ^{11} $ Hz) \cite{Gruner1988}. The bandwidth of our signal is restricted below 8 Hz by the low pass filter of the lock-in amplifier together with a high roll-off digital filter. The lower end of the frequency window is determined by the total time it takes to finish one trace. The background noise, which was simultaneously measured by the out-of-phase component of the lock-in amplifier, consists of mainly thermal noise and does not play a role in the results to be presented below. An ac current was sent through the sample from contacts 1 and 5 to the ground contact 3. The magnitude of the current was defined by the two low noise decade resistors (R1 and R2). When the bridge is balanced, the lock-in amplifier only measures the residual voltage fluctuations, which was recorded as a function of time across the contacts 2 and 4. Contact resistance does not play a role in this configuration. Fig. \ref{Figure1.png} (d) shows the time dependent resistance fluctuations in the CDW pinned state at three different temperatures around T$_{P2}$ (= 59 K). A clear temperature dependence of the magnitude of the resistance fluctuations can be seen. At 68 K, the resistance fluctuation $\Delta$R/R is in the order of 1/10000. As the temperature is reduced below to 42 K, $\Delta$R/R increases by two orders in magnitude. The normalized power spectral density (PSD) of the resistance S$_{R}$/$R^{2}$(f), which represents the magnitude of the resistance fluctuation in the frequency domain, was calculated for these three traces using the Welch's periodogram
method \cite{Welch1967} and is plotted in Fig. \ref{Figure1.png} (e). The PSD traces at all temperatures follow a 1/f$ ^{\alpha} $ behavior with a temperature dependent $ \alpha $ varying around 1. In the rest of this letter, we will present our studies of the temperature and electric field dependence of the resistance fluctuations, focusing on the pinned regime of the CDW.

\begin{figure}
\includegraphics[scale=0.8]{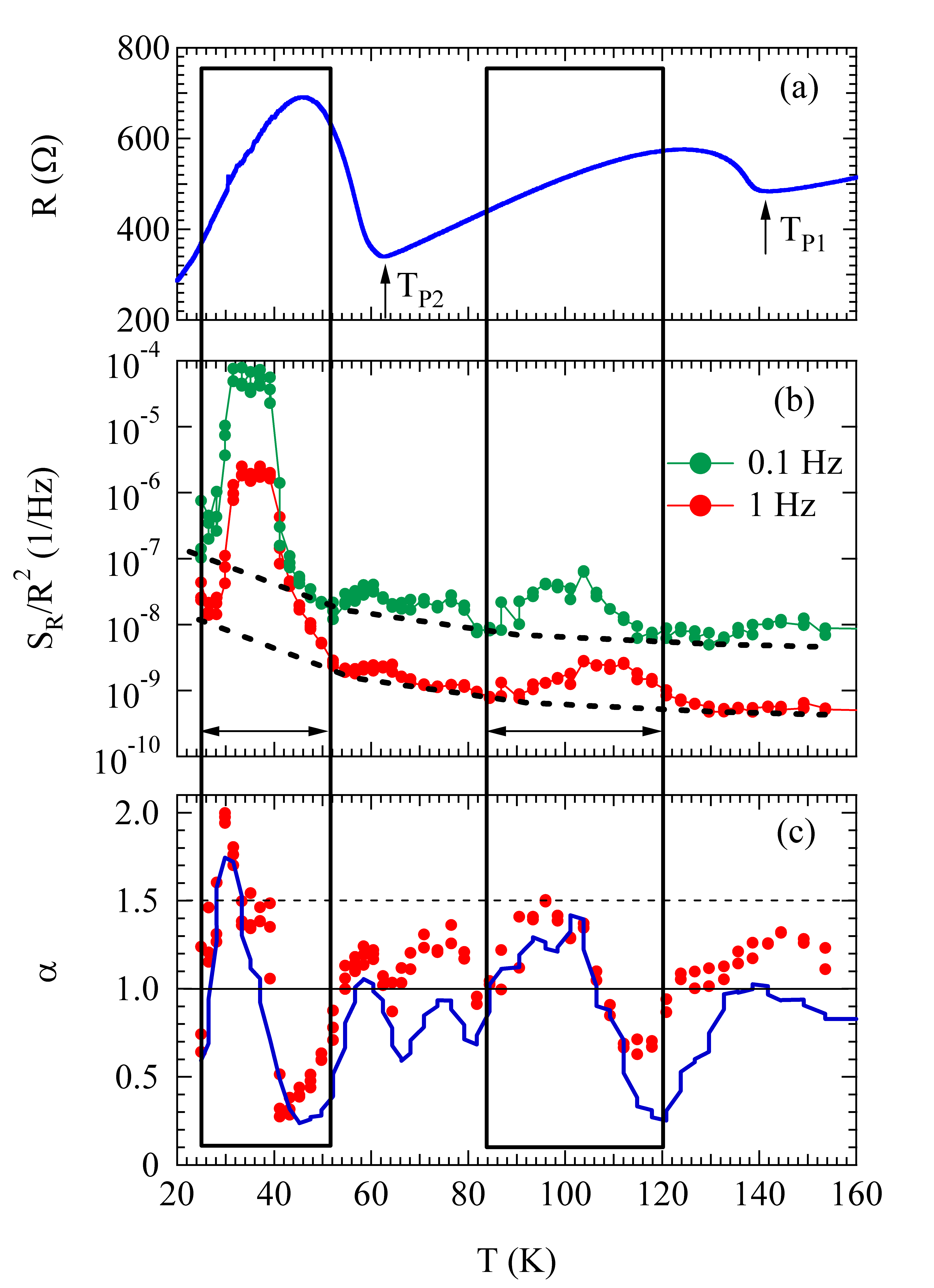}
\caption{(a) R vs. T trace between 20 K and 160 K for device B. (b) S$_{R}$/R$^{2}$ at 1 Hz (red) and 0.1 Hz (green) are plotted as a function of temperature in the same range. The bias current is 1 $\mu$A. The dashed line indicates the general envelope of the noise. The boxed areas marks the temperature range where large excess noise peaks appear. (c) The frequency exponent $ \alpha $ (red) as a function of temperature. The blue trace is the numerical fit to the Dutta-Horn relation \cite{Dutta1981}.}
\label{Figure2.png}	
\end{figure}

In Fig. \ref{Figure2.png} (a), the R(T) of Device B is shown between 20 K and 160 K for comparison. In Fig. \ref{Figure2.png} (b), S$_{R}$/$R^{2}$(1 Hz) and S$_{R}$/$R^{2}$(0.1 Hz) measured in the zero-bias limit (=1 $ \mu $A, 10 times smaller than the depinning current at all temperatures) is plotted in the same temperature range. In all the samples, the magnitude of the resistance noise consists of two parts: a smooth varying background as indicated by the dashed line, and excess noise peaks in the boxed regions. The position and magnitude of the excess noise peaks could vary from sample to sample but are repeatable within the same sample, suggesting a close relation between these excess noise peaks and dynamics of the periodic lattice as determined by imperfections. A comparison of the two traces measured at 0.1 Hz and 1 Hz (in Fig. \ref{Figure2.png} (b)) reveals that, the peak positions shift towards higher temperatures when measured with higher frequencies, suggesting a thermal origin for these peaks. We thus posit that a likely origin for the remarkable discontinuous noise spikes lies in thermally assisted phase slips in the pinned CDW state \cite{Maher1991,DiCarlo1993}. Since the longitudinal dimension of our sample is 5 $\mu$m, the strain on the CDW is expected to be appreciable along the length of the sample, inducing phase slip events \cite{Lemay1998}. Moreover, the typical lateral dimensions of our samples are a few hundred nanometres, which are smaller than the lateral coherence length in pure NbSe$_{3}$ (a few microns) \cite{McCarten1989}. Thus, our samples are expected to behave as a 1D CDW system, in which phase slip events could induce changes in the quasiparticle concentration in the sample \cite{Maher1991,McCarten1992}. In the pinned regime, these phase slip events can have long relaxation times, resulting in metastable states. When the sample dimensions are relatively small, conductance could exhibit discrete jumps as single phase slip events become important \cite{Zybtsev2010, Zybtsev2011}. This suggests that the charge trapping-detrapping model used to understand noise in semiconductors may be relevant. The noise in semiconductors has been shown to follow a smeared activated behavior, as described by Dutta and Horn \cite{Dutta1981,Black1983}. We fit our data with the Dutta-Horn relation 
$-{\dfrac{\partial lnS(f,T)}{\partial lnf}} = 1 + {\dfrac{1}{ln(f_{0}/f)}}\lbrace{\dfrac{\partial lnS(f,T)}{\partial lnT}} - 1\rbrace $ at 1 Hz, and the attempt frequency $ f_{0} $ was chosen to be $ 10^{8} $ Hz. This value is much smaller than the typical phonon frequency $ 10^{14} $ in semiconductors \cite{Dutta1981}, suggesting the phase slip process may be a much slower process. The fit is shown by the blue trace in Fig. \ref{Figure2.png} (c) and it captures the general behaviour of $ \alpha $(T) very well. However, further examination of the D-H fit seems to suggest that the fit to the data is better in the temperature range within the boxed area. Within the boxed areas (25 K - 52 K and 84 K - 120 K), the correlation between $ \alpha $(T) and the Dutta-Horn fit is about 0.9. While outside the boxed areas (52 K - 84 K and 120 K - 155 K), the correlation between $ \alpha $(T) and its Dutta-Horn fit is reduced by half. This seems to suggest that when excess noise peaks are absent the thermally activated behavior alone cannot account for the resistance fluctuations and other effects, such as the mobility fluctuation, may become important \cite{Zybtsev2010,Zybtsev2011}. 

This idea is further supported by the behavior of the general envelope of the noise, as indicated by the dashed line in Fig. \ref{Figure2.png} (b). The general envelope of the noise increases exponentially as the temperature is lowered in all the samples we measured, and it can be fitted by the equation S$_{R}$/R$^{2}$(T) = $\Gamma e^{-k_{B}T/{E_{0}}} $, where $ \Gamma $ is in the order of $ 10^{8}$ Hz$^{-1}$ and $ E_{0} $ is found to be around 1.5 mV. This equation describes an opposite trend of the temperature dependence of the resistance noise compared to the thermally activated behavior described by the Arrhenius equation, as normally seen in semiconductors and metals \cite{RMPv60p537,COISSMSv6p67}. Since the charge carrier density fluctuation due to the phase slippage is thermally activated, it cannot account for this exponential increase of the resistance noise with the lowering of the temperature suggesting the role of mobility fluctuation \cite{Zybtsev2010,Zybtsev2011}. When the sample size is reduced, local defect geometry becomes increasingly important in determining the mobility of the quasiparticles. Especially in high quality single crystal samples, the mean free path could be as large as tens of microns \cite{McCarten1992}, and increases as the temperature is lowered. So any change in the local defect geometry, induced by either an ambient temperature fluctuation or a non-uniform current density, could induce fluctuations in the mobility and resultant noise behavior seen in Fig 2 (b). The role of mobility fluctuation in determining resistance noise has not yet been clearly understood and further experimental and theoretical studies are certainly needed.

\begin{figure*}
\includegraphics[width=6in,height=5in]{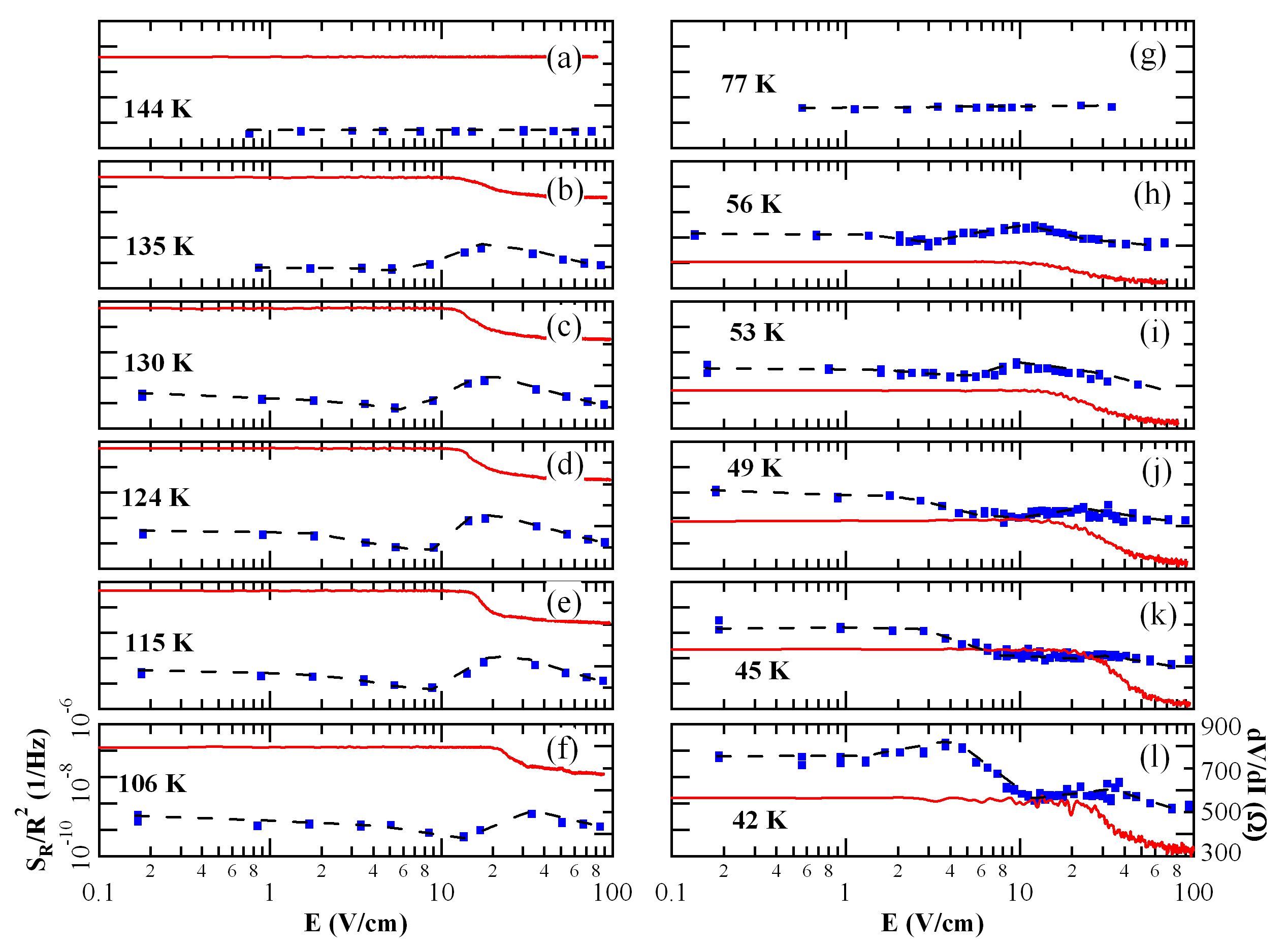}
\caption{(a)-(f) S$_{R}$/R$^{2}$ at 1 Hz (blue dots), as indicated by the left axis, is plotted as a function of electric field for selective temperatures around T$_{P1}$, for device C. The black dashed lines are guides to eye. In the same plots, the differential resistance dV/dI (red curve) are plotted using right axis. At temperatures below the Peierls transition temperatuer (T$_{P1}$ = 142 K for this device), threshold behavior in dV/dI can be seen as a function of electric field. (g)-(l) S$_{R}$/R$^{2}$ at 1 Hz (blue dots, left axis) and dV/dI (red curve, right axis) as a function of electric field for selective temperatures around T$_{P2}$ (= 62 K for this device) are shown for device B. Qualitatively similar features were observed in all the measured samples.}
\label{Figure3.png}
\end{figure*}

Next, we studied the electric field dependence of the noise in our NbSe$_{3}$ nanoribbon samples, both in the pinned state (zero-bias limit) and across the electric field-induced depinning state of the CDW. Before discussing the results in our NbSe$_{3}$ nanoribbon samples, we want to recall some of the previous studies of noise measurements in bulk CDW materials. Noise in CDW materials has been studied extensively and the universal observation was an onset of large noise when the electric field exceeds a threshold (E$ _{T} $) and depins the CDW from pinning centers \cite{Bhattacharya1985,Gruner1988,Maher1991,Bloom1993}. Below E$ _{T} $, the magnitude of the resistance fluctuation was very small and often buried within the background noise. By carefully selecting nanoscale samples and extending the measurements to a lower frequency window, we observed a distinctly different noise behavior in the pinned regime for the first time. In Fig. \ref{Figure3.png}, S$_{R}$/$R^{2}$(1Hz) as a function of electric field (blue dots, left axis) is shown at selected temperatures across T$_{P1}$ for Device C, and across T$_{P2}$ for Device B. In the same graphs, the differential resistance (dV/dI) curves as a function of electric field at the same temperatures were also shown (red solid curve, right axis) to mark the threshold behavior.

In Fig. \ref{Figure3.png} (a), the temperature is 144 K, which is above the Peierls transition temperature (T$_{P1}$ = 142 K). The system behaves as a metal at this temperature and  both S$_{R}$/$R^{2}$(1 Hz) and dV/dI are independent of the applied electric field. As the temperature is reduced below T$_{P1}$, at 135 K in Fig. \ref{Figure3.png} (b), a threshold behavior is seen in the electric field dependence of dV/dI. Close to the threshold electric field, an onset of a large noise was observed, consistent with earlier studies \cite{Bhattacharya1985,Gruner1988,Maher1991,Bloom1993}. The noise in the pinned regime is about one order smaller than the peak value around the depinning field. As the temperature is reduced further, as can be seen in Fig. \ref{Figure3.png} (c)-(f), the onset of a large noise close to the depinning threshold is observed without much change in magnitude. However, the noise in the pinned regime of CDW developed surprising features. A few observations can be made from Fig. \ref{Figure3.png} (c)-(f). First, the noise magnitude at zero bias limit is comparable to the noise magnitude at the depinning regime, while in previous studies, no measurable noise was reported in this pinned regime \cite{Bhattacharya1985,Bloom1993}. Second, the noise in this pinned regime often has a non-monotonic feature in this temperature range (130 K - 106 K) for this sample. Namely, it decreases sharply as the electric field is approaching the threshold, reaching a minimum before the onset of the CDW sliding. Third, the noise at 1 Hz and 0.1 Hz (not shown) generally follow similar trend though minor differences can be observed. 

We performed similar measurements around T$_{P2}$ in another sample (Device B), and the results are shown in Fig. \ref{Figure3.png} (g)-(l). At 77 K (Fig. \ref{Figure3.png} (g)), the system behaves as a metal, and S$_{R}$/$R^{2}$ is independent of electric field. As the temperature is lowered below T$_{P2}$ (62 K), as shown in Fig. \ref{Figure3.png} (h)-(l), the threshold behavior was observed in both noise and dV/dI curves, similar to the behavior observed around T$_{P1}$. However, a clear difference between the behavior of the noise around T$_{P1}$ and T$_{P2}$ can immediately be seen. As shown in Fig. \ref{Figure3.png} (h)-(l), below T$_{P2}$, the magnitude of the noise in the pinned regime grew much faster with decreasing temperature than that close to the depinning threshold. At some temperatures, the depinning noise was even buried within the tail of the large noise in the pinned regime, as shown in Fig. \ref{Figure3.png} (k) and (l). In comparison, in Fig. \ref{Figure3.png} (b)-(f), the change in the noise magnitude in the pinned regime when reducing the temperature was much smaller. We want to note that the noise peak magnitude close to the depinning threshold does not change much as a function of temperature. In all nine samples we measured, it is always $10^{-9}$ - $10^{-8}$ Hz$^{-1}$ around T$_{P1}$, and $10^{-8}$ - $10^{-7}$ Hz$^{-1}$ around T$_{P2}$. Thus, it can be stated different mechanisms are responsible for the origin of the noise in the pinned regime and the sliding regime. Another observation can be made from Fig. \ref{Figure3.png} (l). In the pinned regime, the noise reaches a peak value around 4 V/cm. This field is much smaller than the depinning threshold (20 V/cm). When the electric field is increased further from 4 V/cm to 10 V/cm, the noise magnitude drops sharply. This noise peak feature in the pinned regime is a very sensitive function of temperature, electric field and frequency. But one feature is always present: the noise peak feature in the pinned regime is always suppressed when threshold electric field is approached. These noise peaks in the pinned regime are reminiscent of the discrete fluctuators that were observed in the CDW sliding regime \cite{Bloom1994}. 

Combining the temperature dependence and the electric field dependence studies, a clearer picture about the origin of the resistance noise in the CDW pinned state can be obtained for the first time. The non-monotonic behavior of the noise with increasing bias current in the pinned regime was only seen when it is measured within the temperature range where the excess noise peaks are seen (boxed area in Fig. \ref{Figure2.png} (b)). These excess noise peaks are thought to be induced by thermally activated phase slip events, which can lead to quasiparticle density fluctuations. Only when the sample size is small individual phase slips and the interference between a finite number of phase slip events across the cross section of the naonoribbon become important, and the non-monotonic behavior of the noise with increasing bias current in the pinned regime could be observed in our nanoribbon samples. 
When the electric field approaches the threshold, CDWs are progressively depinned and the CDW strain is largely suppressed rendering a reduction of the phase slip events, which is responsible for the local minimum in noise magnitude just before the onset of CDW sliding. This non-monotonicity in the noise magnitude thus may denote a crossover for the origin of the noise from phase slip events dominated within the pinned regime to a combination of phase slips, pinning strength fluctuations and non-equilibrium effects of the flowing CDW \cite{Zybtsev2010,Maher1991,Bhattacharya1989,Bloom1994}.

In summary, we have provided the first systematic study of the 1/f noise in the CDW pinned regime in NbSe3 nanoribbons. Some hitherto unobserved features are noted in the 1/f noise magnitude as the threshold electric field to depin the CDW state is approached. Within the temperature range where excess resistance noise peaks are seen (Fig. \ref{Figure2.png} (b)), the 1/f noise behavior indeed showed signatures of thermally activated phase slip events. However, a non-thermal behavior, manifested as an exponential increase of the noise magnitude as the temperature is lowered, was also observed across the rest of the temperature range. A model involving mobility fluctuation could explain the general envelope of the noise as a function of temperature.

We are grateful to Dr. Arindam Ghosh (IISc, Bangalore) and his students for helpful discussions about the 1/f noise measurement and digital signal processing techniques. This work was supported by the National Science Foundation under DMR 0847324;  P. M, K. F. and S.B. acknowledge support from the National Science Foundation under IIP 1311837.

\bibliographystyle{apsrev}
\bibliography{CDWpaper}
\end{document}